\documentclass[10pt,aps,prl,superscriptaddress,floatfix,dvipsnames,twocolumn,nobibnotes]{revtex4-2}

\usepackage{graphicx,xcolor,url,hyperref,dcolumn,bm,multirow,amsmath,amssymb,amsfonts,subfigure,soul,textcomp}
\usepackage[version=3]{mhchem}
\usepackage[normalem]{ulem}
\newcommand{\m}{\mathbf}
\newcommand{\coso}{Cu$_2$OSeO$_3$}
\newcommand{\tc}{$T_{\textrm{C}}$}

\usepackage[scaled]{helvet}
\usepackage[T1]{fontenc}
\hypersetup{
	hidelinks, colorlinks  = true, linkcolor   = blue, citecolor   = Red, urlcolor = blue
}

\begin{document}
	
	\title{A Surface-confined Spiral State With The Double Period \\ In The Cubic Chiral Helimagnet \ce{Cu2OSeO3}}
	
	\author{Priya R. Baral}
	\email{priya.baral@epfl.ch}
	\affiliation{Department of Applied Physics and Quantum-Phase Electronics Center (QPEC), The University of Tokyo, Bunkyo-ku, Tokyo 113-8656, Japan}
	\affiliation{PSI Center for Neutron and Muon Sciences, Paul Scherrer Institute (PSI), CH-5232 Villigen, Switzerland}
	\affiliation{Institute of Physics, \'Ecole Polytechnique F\'ed\'erale de Lausanne (EPFL), CH-1015 Lausanne, Switzerland}

	\author{Oleg I. Utesov}	\affiliation{Center for Theoretical Physics of Complex Systems, Institute for Basic Science, Daejeon 34126, Republic of Korea}
	
	\author{Samuel H. Moody}
	\affiliation{PSI Center for Neutron and Muon Sciences, Paul Scherrer Institute (PSI), CH-5232 Villigen, Switzerland}
	
	\author{Matthew T. Littlehales}
	\affiliation{Department of Physics, Durham University, Durham DH1 3LE, United Kingdom}
	\affiliation{ISIS Neutron and Muon Source, Rutherford Appleton Laboratory, Didcot, OX11 0QX, United Kingdom}
	
	\author{Pierluigi Gargiani}
	\affiliation{ALBA Synchrotron Light Source, E-08290 Cerdanyola del Vall\'es, Barcelona Spain}
	
	\author{Manuel Valvidares}
	\affiliation{ALBA Synchrotron Light Source, E-08290 Cerdanyola del Vall\'es, Barcelona Spain}
	
	\author{Robert Cubitt}
	\affiliation{Institut Laue–Langevin, 71 avenue des Martyrs, CS 20156, Grenoble, 38042 Cedex 9, France}
	
	\author{Nina-Juliane Steinke}
	\affiliation{Institut Laue–Langevin, 71 avenue des Martyrs, CS 20156, Grenoble, 38042 Cedex 9, France}
	
	\author{Chen Luo}
	\affiliation{Helmholtz-Zentrum Berlin f\"ur Materialien und Energie, D-12489 Berlin, Germany}
	
	\author{Florin Radu}
	\affiliation{Helmholtz-Zentrum Berlin f\"ur Materialien und Energie, D-12489 Berlin, Germany}
	
	\author{Arnaud Magrez}
	\affiliation{Institute of Physics, \'Ecole Polytechnique F\'ed\'erale de Lausanne (EPFL), CH-1015 Lausanne, Switzerland}
	
	\author{Jonathan S. White}
	\affiliation{PSI Center for Neutron and Muon Sciences, Paul Scherrer Institute (PSI), CH-5232 Villigen, Switzerland}
	
	\author{Victor Ukleev}
	\email{victor.ukleev@helmholtz-berlin.de}
	\affiliation{Helmholtz-Zentrum Berlin f\"ur Materialien und Energie, D-12489 Berlin, Germany}

	\keywords{magnetic anisotropy, anisotropic exchange, helimagnetism, chiral magnet, skyrmions}
	
	\date{\today}
	
	\clearpage
	
	\newpage
	
	\flushbottom
	\maketitle

	\section{Abstract}
	\textbf{The chiral magnetoelectric insulator \coso\,hosts a rich and anisotropic magnetic phase diagram that includes helical, conical, field-polarised, tilted conical, and skyrmion lattice phases. Using resonant elastic x-ray scattering (REXS), we uncover a new spiral state confined to the surface of \coso. This surface-confined spiral state (SSS) displays a real-space pitch of $\sim$120 nm, which remarkably is twice the length of the incommensurate structures observed to-date in \coso. The SSS phase emerges at temperatures below 30~K when the magnetic field is applied between $3^\circ$ to $18^\circ$ away from the $\langle\text{110}\rangle$ crystallographic axes. Its surface localisation is demonstrated through a combination of REXS in reflection and transmission geometries, with complementary small-angle neutron scattering measurements suggesting its absence from the bulk. We attribute the stabilisation of the SSS to competing anisotropic interactions at the crystal surface. The discovery of a robust, surface-confined spiral paves the way for engineering energy-efficient, nanoscale spin-texture platforms for next-generation devices.}

	\section{Introduction}
	
\begin{figure*}[htb!]
		\begin{center}
			\includegraphics[width=1\linewidth]{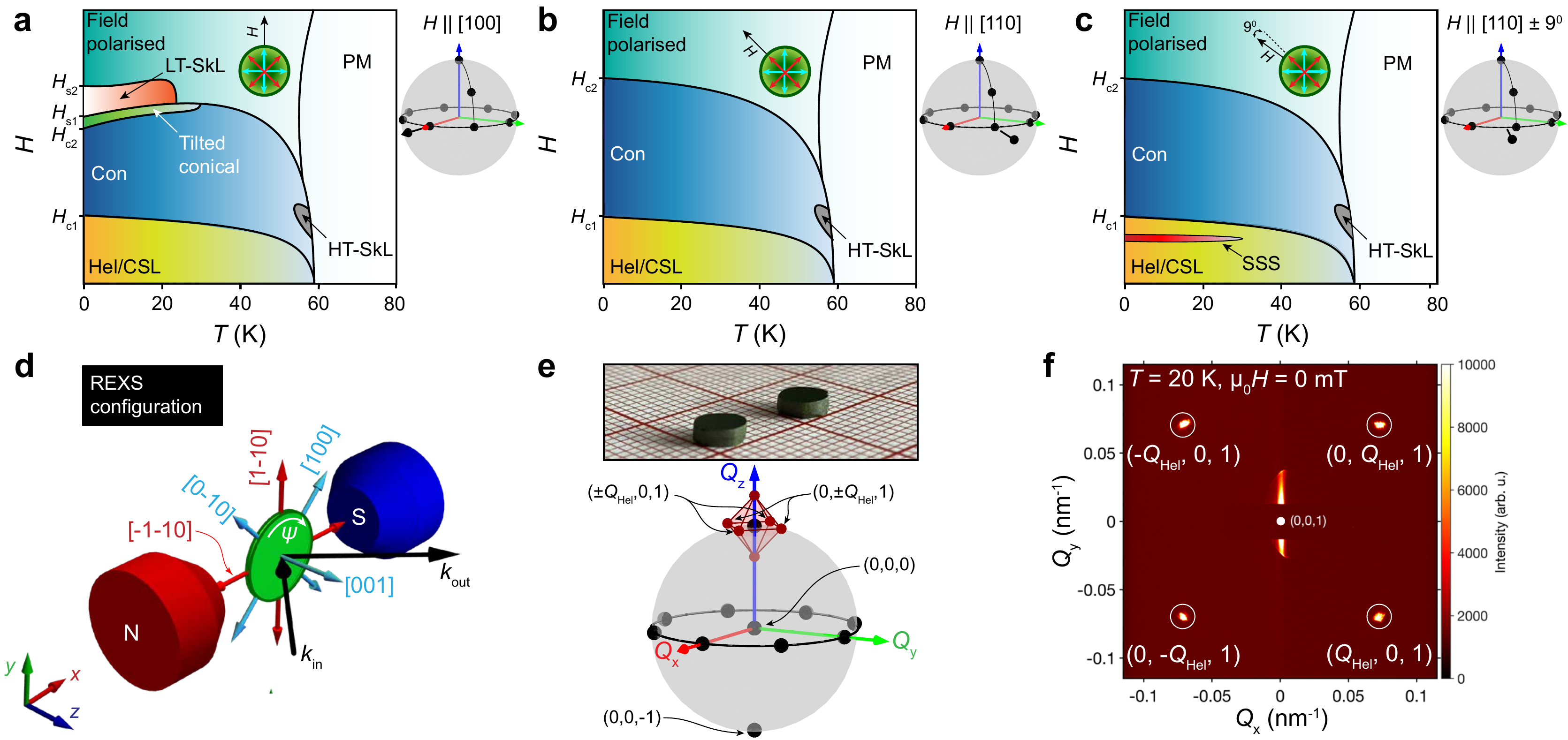}
			\caption{\textbf{Resonant Elastic X-ray Scattering (REXS) and helimagnetism in \ce{Cu2OSeO3}.} Schematic phase diagram of \coso\,while the magnetic field is applied along \textbf{(a)} [100], \textbf{(b)} [110], and \textbf{(c)} 9 degrees away from [110] crystallographic axis, highlighting the appearance of the surface-confined spiral state (SSS). The green-filled circle represents the disc sample used in the current study. Cyan and red arrows represent the in-plane (100) and (110) domains, respectively, while [001] points in the out-of-plane direction. The SSS appears below the conical (Con) $\leftrightarrow$ helical (Hel) transition. Note that low-temperature skyrmion lattice (LT-SkL) phase is stabilised only along $\langle\text{100}\rangle$, in contrast to more isotropic high-temperature skyrmion lattice (HT-SkL) phase. The schematics shown in Panels-\textbf{a} and \textbf{b} are drawn according to the experimental phase diagrams reported in Ref.~\cite{halder2018thermodynamic}. \textbf{(d)} Experimental configuration for our REXS experiment at BL-29 BOREAS beamline. The $x$-direction is always fixed along the magnetic field. The disc-shaped \coso\,crystal can be rotated perpendicular to [001] axis, giving access to four in-plane (100) and (110) domains. \textbf{(e)} Top: optical image of the disc-shaped single crystal used for scattering studies reported in this work. Bottom: reciprocal space of \coso\, in magnetically ordered state at zero field. The six red spheres surrounding (001) reflection are the results of three $\mathbf{Q}_{\mathrm{Hel}}$ domains demonstrating local octahedral ($O_{h}$) symmetry. For simplicity, $\mathbf{Q}_{\mathrm{Hel}}$ peaks around the other (100), as well as (000) reflections, have been omitted. \textbf{(f)} REXS pattern collected for \coso\, crystal at $T$ = 20~K in absence of external magnetic field. As labeled in the figure, the four diffraction peaks are a consequence of the four in-plane helical domains as shown in panel-\textbf{e}.}
			\label{fig1}
		\end{center}
	\end{figure*}

	\noindent Chiral magnetic textures are promising building blocks for quantum technologies. Helical and conical spin textures \cite{yokouchi2020emergent,kitaori2021emergent,kitaori2023doping}, chiral soliton lattices (CSL)\cite{togawa2013interlayer}, magnetic hedgehogs \cite{kanazawa2011large} and topologically-nontrivial skyrmions \cite{bogdanov1994thermodynamically,muhlbauer2009skyrmion,fert2017magnetic,nayak2017magnetic} promote emergent electromagnetic phenomena that can be crucial for developing the next-generation dissipationless electronics \cite{tokura2017emergent}. Chiral cubic magnets provide the fertile ground to nucleate such spin textures due to the competition among Heisenberg exchange interaction, antisymmetric Dzyaloshinskii-Moriya interaction (DMI), anisotropic exchange interaction (AEI), and cubic anisotropy \cite{bak1980theory,nakanishi1980origin,maleyev2006cubic}. The energy density of such cubic helimagnets within the Bak-Jensen~\cite{bak1980theory} model reads as:
	
	\begin{equation}
		\begin{split}
			\varepsilon = \frac{J}{2}(\nabla \m{M})^2 - D \m{M} \cdot &\nabla \times \m{M} + \frac{F_\textrm{AEI}}{2} \sum_{\nu=x,y,z} \left(  \partial_\nu \m{M}_\nu \right)^2  \\
			& +	K \sum_{\nu=x,y,z} M^4_\nu - \m{H} \cdot \m{M}
		\end{split}
		\label{TE}
	\end{equation}
	
	where, the first term corresponds to exchange interaction, the second to DMI, the third to AEI, the fourth to cubic anisotropy, and the final to the Zeeman energy. By tuning the fragile balance between these interactions, and by varying external parameters, such as temperature, magnetic field, and pressure, one can achieve a plethora of modulated magnetic states, such as helical and conical spin textures \cite{nakanishi1980origin,lebech1989magnetic}, triangular and square skyrmion lattices (SkL) \cite{muhlbauer2009skyrmion,seki2012observation,nakajima2017skyrmion,karube2020metastable,takagi2020particle}, disordered skyrmions \cite{chacon2018observation,karube2018disordered,bannenberg2019multiple}, three-dimensional magnetic hedgehog lattice \cite{kanazawa2011large}, CSL \cite{okamura2017emergence,nakajima2018uniaxial,ukleev2020metastable}, elongated skyrmions \cite{morikawa2017deformation,nagase2019smectic,aqeel2021microwave}, and their combinations. More specifically in \ce{Cu2OSeO3}, member of the skyrmions hosting $P2_13$ magnets, the phase diagram is strongly affected by the intrinsic balancing between anisotropic magnetic interactions with the external magnetic field \cite{qian2018new,chacon2018observation,bannenberg2019multiple,leonov2020field,moody2021experimental,leonov2023reorientation,baral2023direct,marchiori2024imaging,mehboodi2024observation}. Thus, in case of $\mu_0 H || [001]$, low-temperature disordered skyrmion (LT-SkL) and tilted conical (TC) phases emerge in addition to the well-defined helical, conical and SkL (HT-SkL or the so-called $\mathcal{A}$-phase just below the Curie temperature \tc\,$\simeq$ 58\,K) phases \cite{halder2018thermodynamic} (shown schematically in Figs. \ref{fig1}\textbf{a},\textbf{b}).
	
	Furthermore, the effects of the geometrical confinement in nanostructured chiral magnets, including \ce{Cu2OSeO3}, often lead to modifications of their magnetic phase diagrams and emergence of new magnetic states \cite{yu2011near,zhao2016direct,jin2017control,pathak2021geometrically,niitsu2022geometrically,baral2022tuning}. Additional interactions arising due to surface-symmetry breaking in a chiral magnet may promote unique magnetic textures that are not present in the bulk material, yet may exist in technologically relevant low-dimensional surface settings \cite{rybakov2015new,rybakov2016new,zhang2018reciprocal,zheng2018experimental,zhang2018direct,zhang2020robust,burn2021periodically,turnbull2022x,xie2023observation,jin2023evolution}.
	
	Here, we report the discovery of a previously unknown magnetic structure in \coso~with a doubled period in comparison with the conventional bulk helix at $T<30$\,K in the narrow window of magnetic fields applied slightly away from the $\langle\text{110}\rangle$ directions (Fig. \ref{fig1}\textbf{c}). By combining REXS in both reflection and transmission geometries with small-angle neutron scattering probes, we show that it is both metastable and confined near the surface of \coso~and call it surface-confined spiral state (SSS). Our experimental data support the notion that SSS is stabilised as a result of a subtle interplay among various anisotropic interactions. In particular, the axial surface anisotropy should be crucial for its stability. 
	
	\section{Results}
	
	\noindent \textit{\textbf{REXS at low-temperature in chiral cubic \ce{Cu2OSeO3}.}} REXS in the Bragg geometry has been used to observe magnetic satellites of the anomalous (001) reflection \cite{langner2014coupled,zhang2017direct} (Figs.~\ref{fig1}\textbf{d}, and bottom panel of \textbf{e}). This reflection is accessible at the Cu $L_3$ edge thanks to the large unit cell of \coso. Recently, REXS has been proven to be a very effective probe to uncover subtle features in the magnetism of \coso~\cite{zhang2018reciprocal,zhang2020robust,ukleev2022chiral}. The surface sensitivity of REXS provides us with an added layer of information, which remains invisible to other scattering probes, such as bulk-sensitive neutron scattering. As calculated by Zhang \textit{et al}., at the incident angle of 48$^\circ$ corresponding to the Bragg condition of (001) diffraction peak, the penetration depth of x-rays varies from $\sim30$\,nm to $\sim100$\,nm in \coso~depending on energy across the resonant $L_3$ edge \cite{zhang2018reciprocal}. This makes REXS an ultimate surface-sensitive technique. 
	
	\begin{figure*}
		\begin{center}
			\includegraphics[width=1\linewidth]{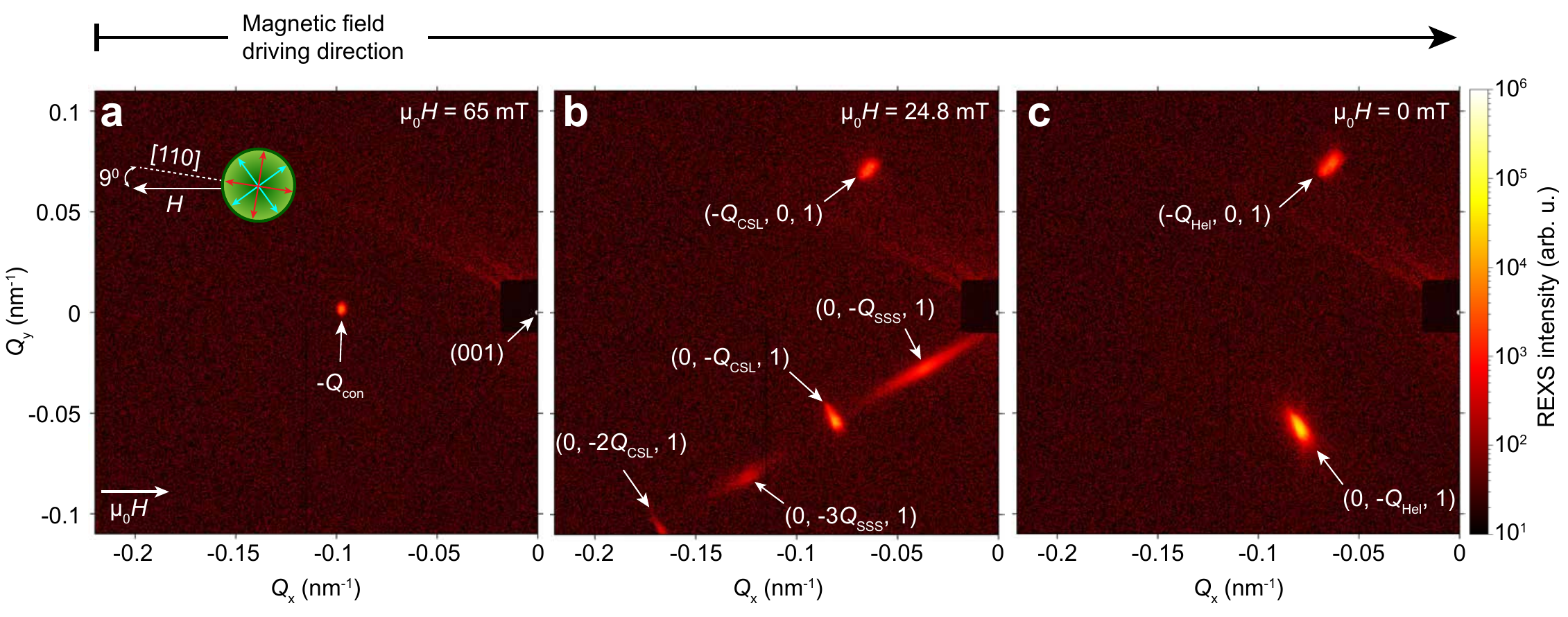}
			\caption{\textbf{Scattering signature of the surface-confined spiral state (SSS) near the conical$\leftrightarrow$helical phase transition.} REXS patterns measured at $T=18$\,K for the field tilted away from the $\langle\text{110}\rangle$ by 9$^\circ$, corresponding to \textbf{(a)} conical, \textbf{(b)} distorted helicoids together with surface-confined spiral state and \textbf{(c)} proper screw spiral states. Additional peaks appear due to the SSS with a doubled helical pitch. Note that we do not observe the SSS while ramping the magnetic field in the opposite direction, that is, while accessing the phases in the following sequence: helical$\rightarrow$conical$\rightarrow$field-polarised state. Applied magnetic field direction is indicated by the horizontal white arrow in Panel~\textbf{a}. The colorbar scale is the same for all three detector maps. The inset in Panel-\textbf{a} shows the relative orientation of applied magnetic field with respect to the $\langle\text{110}\rangle$ crystallographic axis.}
			\label{fig2}
		\end{center}
	\end{figure*}
	
	We used one \coso~single crystal, grown by chemical vapor transport (CVT), for all the main results reported in this article. Further information can be found in Methods. In order to reduce the demagnetisation effects, the single crystal was carefully cut and polished to have a disc shape, as shown in top panel of Fig.~\ref{fig1}\textbf{e}. The zero-field magnetic REXS pattern shows up as four clear Bragg peaks on the detector plane, corresponding to the helical state (Fig. \ref{fig1}\textbf{f}). In the subsequent experiment, only a half of the reciprocal space ($-0.25 \leq \mathbf{Q}_x\leq 0$\,nm$^{-1}$) was measured to access the higher harmonics of magnetic satellites in the single detector configuration at a finite magnetic field. At $T=20$\,K, the magnetic field of 100\,mT drives the system into the spin-polarised phase. Reduction of the field results in the gradual emergence of the conical satellite (Fig. \ref{fig2}\textbf{a}). When the field is aligned roughly along  $\langle\text{110}\rangle$ axes, upon further decrease of the magnetic field -- just below 30\,mT -- a first-order transition from conical to the multi-domain helical phase takes place, manifesting two Bragg reflections corresponding to helices slightly tilted away from their preferable $\langle100\rangle$ directions by the remaining magnetic field. Due to the finite magnetic field, the proper-screw structure is deformed, resulting in higher-harmonic reflections, similar to the case of CSL in uniaxial helimagnets \cite{togawa2013interlayer, kishine2015}.
	
\begin{figure*}
		\begin{center}
			\includegraphics[width=1\linewidth]{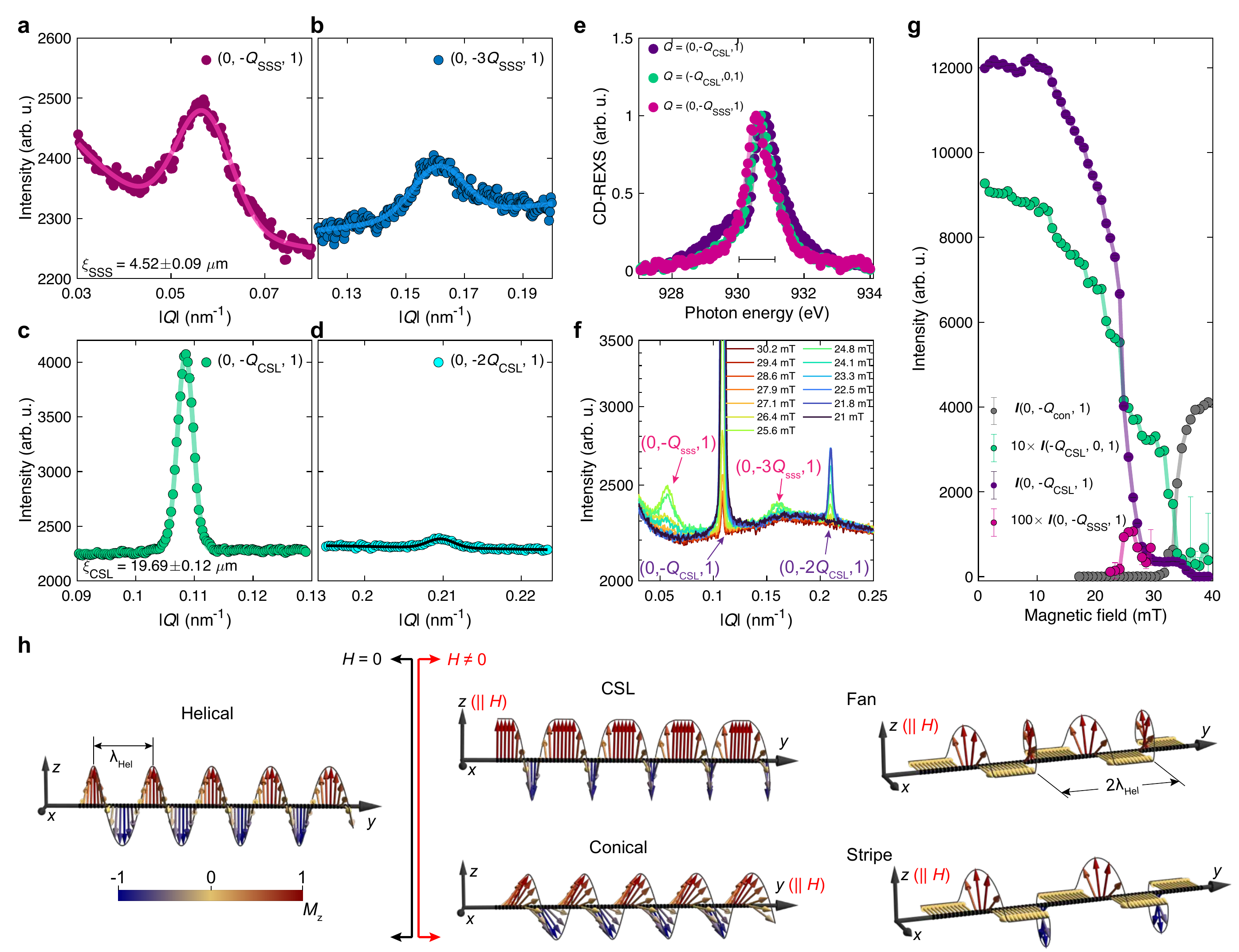}
			\caption{\textbf{Magnetic field-induced evolution of $\mathbf{Q}_\text{SSS}$ and its relationship with other chiral textures.} \textbf{(a)}-\textbf{(d)} The $Q$-dependence of various magnetic satellites and their higher harmonics was extracted from the REXS data and analysed using Eq.~S1 (described in the Supplementary Material). For a direct comparison of scattering intensities, the $y$-scale was kept identical across corresponding harmonics. Panel \textbf{(e)} shows the normalised CD-REXS energy dependence of both CSL and SSS peaks at the Cu $L_3$ resonant edge. Measurements were performed at $T=18$~K. FWHM of a Gaussian fit to the surface-confined spiral data is indicated. \textbf{(f)} Magnetic field dependence of the $Q$-scan along (0,-$Q$,1) direction obtained from the \ce{Cu2OSeO3} disc at $T$ = 18~K. For all scans, the incident photon energy was fixed at 930.8~eV. \textbf{(g)} Integrated intensity as a function of magnetic field for all four fundamental incommensurate reflections measured in our experiment. $I(-Q_\text{CSL},0,1)$ and $I(0,-Q_\text{SSS},1)$ have been scaled up for clarity. The error bars result from the fitting function shown in Eq.~S1. \textbf{(h)} Real-space visualisation of the discussed spin structures throughout the text. Note the different magnetic field direction for conical structure ($\parallel$~\textit{y}) and CSL, fan \& stripe structures ($\parallel$~\textit{z}). The alternating domain wall chirality characteristic of fan structures, which contrasts with the selected chirality of stripe textures, is a crucial distinction between the two magnetic configurations.}
			\label{fig3}
		\end{center}
	\end{figure*}

	\noindent\textit{\textbf{Surface-confined spiral state observed by REXS.}} Further reduction of the field, while being aligned 9$^\circ$ away from $\langle\text{110}\rangle$, leads to unexpected and highly non-trivial results. As evident from Fig. \ref{fig2}\textbf{b}, a new peak appears at the $\mathbf{Q}$ position equal to 1/2 of the helical (or CSL) wavevector, before returning to characteristic multi-domain helical ground state at zero field (see Fig.~\ref{fig2}\textbf{c}). The new peak is clearly broader and less intense than the CSL satellites. As shown in Figures~\ref{fig3}\textbf{a}-\textbf{d}, the SSS correlation length is approximately one-quarter that of the CSL phase and, notably, demonstrates only the 3$^\textrm{rd}$ order of the higher harmonics. Suppression of the second-order peak and emergence of the 3$^\textrm{rd}$ harmonic is a signature of a square-wave magnetic modulation typical to stripe magnetic domains \cite{hellwig2003x,stellhorn2019control}. These additional peaks are only observed in the narrow magnetic field range 22-28\,mT and only at $T<30$\,K and thus are characteristic to the SSS. Importantly, the SSS exhibits extreme sensitivity to the relative orientation between the crystal axes and the applied magnetic field. The SSS stability window spans from $3^\circ$ to $18^\circ$ away from the $\langle\text{110}\rangle$ high-symmetry axes, with maximum scattering intensity observed at $\psi~=9^\circ$. Further details regarding our comprehensive wide azimuthal angle-dependent REXS measurements can be found in the Supplementary Materials  Fig.~S1.
	
	\noindent\textit{\textbf{Spin structure of the novel surface-confined state.}} Chirality of the magnetic texture, and its Bloch or N\'eel-type character can be uniquely identified by means of the circular dichroism in REXS (CD-REXS) technique \cite{zhang2018reciprocal}. We first employed CD-REXS to characterise the dichroic signal in the zero-field helical state, as shown in Fig.~S2. The Friedel pairs exhibit opposite dichroic signals: positive on one side and negative on the other, while near vertical detector axis (along $\langle\text{110}\rangle$) forms the \textit{extinction direction}, where dichroic signal vanishes. Furthermore, Fig. \ref{fig3}\textbf{e} shows the energy dependence of the CD-REXS intensity measured for the two CSL, and the main SSS peak ($\mathbf{Q}_\mathrm{SSS}=\mathbf{Q}_\mathrm{CSL}/2$). The same sign and magnitude of the CD-REXS signal for all three modulations suggest the chiral, Bloch-type nature of the new state.
	
	Having established the Bloch-type chirality of the SSS with its doubled periodicity through CD-REXS measurements, we now address the underlying magnetic structure through a phenomenological description based on the sine-Gordon model for systems with strong axial anisotropy. By incorporating the experimentally observed doubled periodicity as a constraint, we calculated the expected amplitudes of the SSS harmonic components (see Supplementary Materials). Our analysis demonstrates that the second harmonic amplitude is negligible whilst the intensity ratio between the first and third harmonics exhibits excellent agreement with the experimental scattering data. The predicted stripe phase configuration, however, exhibits considerable sensitivity to additional anisotropic interactions. This fragility with respect to perturbations in anisotropy may account for the absence of SSS in the thin transmission REXS lamella, where surface-induced strain and modified interfacial interactions could suppress the delicate balance required for SSS stabilisation.
	
	\section{Discussion}
	
	\noindent The half-order peak position suggests that the spiral periodicity of the emergent phase is doubled compared to the ground-state helical and field-induced CSL and conical states. The phase has a clearly metastable character since it emerges only from the conical phase on the field decrease, and does not emerge from the ground-state helical phase when the field is ramped up. Note, that neither neutron scattering experiments (Supplementary Materials Fig.~S5) nor precision magnetic susceptibility and heat capacity measurements \cite{halder2018thermodynamic} on bulk \coso~crystals could observe this new phase for $\mu_0 H ||\langle110\rangle$. This fact, as well as the broadening of the corresponding $Q_\mathrm{SSS}$ and $3Q_\mathrm{SSS}$ satellites (Fig. \ref{fig3}\textbf{f} and Fig.~S1\textbf{b}) strongly suggests the surface-confined nature of the new spiral state. The metastability of the new surface-confined spiral is clearly manifested in its hysteretic behaviour and its embedding into the CSL phase (Fig. \ref{fig3}\textbf{g}).
	
	The reduced SSS correlation length reflects a combination of geometric and structural factors. First, surface confinement introduces an intrinsic geometric contribution: because the SSS is localised within a thin surface layer ($\sim$30–100~nm), the scattering volume is highly restricted in the out-of-plane direction ($q_z$), and according to the Scherrer equation, this finite thickness inherently broadens the peak in $q_z$. Second, the in-plane ($q_x$) broadening points to surface disorder and pinning effects. As shown theoretically in Ref.~\cite{utesov2015spiral}, disorder in helical magnets induces local distortions in the helix spin rotation angle, creating diffuse scattering that widens the Bragg peak. Given that the SSS is a fragile state stabilised by a delicate balance of surface-modified anisotropies, it is likely more susceptible to pinning by surface defects than the robust bulk CSL. This pinning, combined with the energy cost of maintaining double-periodicity near domain boundaries, ultimately limits the coherent SSS domain size.
	
	Intriguingly, the SSS phase was also not observed in a recent surface-sensitive scanning magnetic force microscopy (MFM) study~\cite{milde2020field}. Apart from the local nature and limited field of view of the MFM probe, we note that the magnetic field in that study was applied precisely along out-of-plane [110] axis, which lies outside the SSS stabilisation window. As we show in the scanning-REXS measurement (Supplementary Materials Fig.~S3), SSS domains are not distributed on the sample surface homogeneously, but with the tendency to form domains of size ca. 500 $\mu$m at the sample edges. Yet, we reiterate that the surface may exhibit a sizable perpendicular net magnetisation which leads to demagnetisation fields that can contribute to a break into domains of the underlying magnetic textures. We emphasise that despite spatial inhomogeneity in SSS intensity across the sample surface, the SSS exhibits systematic spatial correlation with the underlying bulk CSL phase—specifically, higher SSS intensity appears preferentially along $\left[Q = (0,\pm Q_\text{CSL},1)\right]$, that correspond to CSL domains proximal to the applied magnetic field, rather than at arbitrary surface edges or corners. It is important to note that we have successfully observed the SSS from two different crystals of different shapes and independent surface treatment routines. In both samples, the magnetic field tilting angle (between $3^\circ$ and $18^\circ$ away from $\langle\text{110}\rangle$) and field range (ca. 25-30\,mT) needed to stabilise the SSS were similar.
	
	Interestingly, SSS does not show up in the transmission small-angle X-ray scattering (SAXS) experiment on \coso~(see Supplementary Materials Fig.~S4). We believe that the tensile strain in the lamella which provides tangible easy-plane anisotropy destroys the subtle balance among anisotropic interactions and precludes the formation of SSS. Another possible reasoning connected with the existence of the third [001] helical domain in the bulk sample is unlikely according to our estimations, which show that for an in-plane magnetic field this domain would have a larger energy as compared to the other experimentally observable domains. It is worth mentioning that the crystal surface of a lamella may exhibit different structural properties with respect to the surface of a single crystal.
	
	A mechanism similar to the tilted spiral stabilisation in combination with the demagnetisation field at the surface,  may account for the new metastable SSS with the doubled period. Indeed, its emergence only for a narrow angular range of $\psi$ suggests that the balance between cubic and exchange anisotropies is required to achieve the stability of the state. However, the reason why the spiral pitch is doubled, and appears only in the narrow angular range of the applied field, remains unclear at present. In principle, the helical state in cubic chiral magnets is well described by the Bak-Jensen model Eq.~\ref{TE} that determines the spiral periodicity $\lambda\simeq J/D$ as the main term, and the AEI responsible for the small anisotropic corrections, as discussed in the previous studies \cite{ukleev2021signature,moody2021experimental,baral2023direct}. Hence, the doubling of the helical pitch at the surface is unexpected within this phenomenological model.
	
	\begin{figure}[htb!]
		\includegraphics[width=1.0\columnwidth]{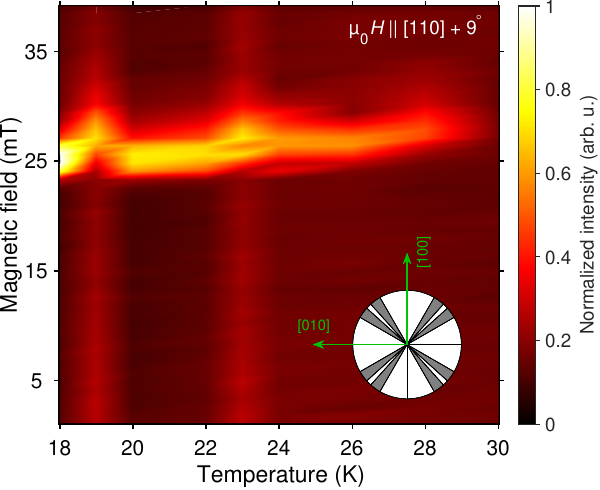}
		\caption{\textbf{\textit{H-T} phase diagram of the surface-confined spiral state.} Magnetic phase diagram for the surface-confined spiral state, as obtained from our REXS data on the disc-shaped sample. For simplicity, intensity stemming from the reflections arising due to conical, CSL as well as Helical phases have not been shown.}
		\label{fig4}
	\end{figure}
	
	Moreover, the SSS is clearly different from the three-dimensional surface spirals stabilised by the demagnetisation field (also termed \textit{stacked spin spirals}) predicted by Rybakov \textit{et al.}~\cite{rybakov2016new}, and recently observed in thin plates of FeGe~\cite{turnbull2022x}. The main differences are: the anisotropic nature of the SSS, and the exact, and field-independent 2$\lambda$ periodicity. The theoretical and experimentally measured periods of stacked spirals in FeGe, $\lambda_\textrm{ss}$ are field-dependent and have a period of $1.1 \lambda \le \lambda_\textrm{ss} \le 1.8 \lambda$ \cite{rybakov2016new}. Similar considerations also apply to the distorted tilted conical stripes observed at the surface of \coso~that change their period as a function of the applied field \cite{marchiori2024imaging,mehboodi2024observation}.
	
	Recently, stability of square-wave-like spin textures, so-called stripe and fan states was shown in the uniaxial chiral magnet MnNb$_3$S$_6$ as an outcome of the competition between DMI, easy-plane anisotropy and dipolar interactions \cite{karna2021annihilation,osorio2023chiral}.  Moreover, the modulation period of these textures is about twice longer than the pitch of the CSL \cite{osorio2023chiral}. Schematic illustrations of the helical, conical, CSL, fan, and stripe magnetic textures are given in Fig. \ref{fig3}\textbf{h}. It is important to note, that the fan state is comprised of Bloch-type domain walls with alternating chirality \cite{moriya1982evidence}. Nevertheless, the fan texture has been experimentally observed \cite{karna2021annihilation} and reproduced by means of micromagnetic simulations \cite{osorio2023chiral} in the chiral-lattice MnNb$_3$S$_6$ as a result of the interplay between the interactions mentioned earlier.
	
	As is well established for a number of cubic chiral magnets, including \ce{Cu2OSeO3}, uniaxial anisotropy emerges in these systems as a consequence of mechanical strain~\cite{nakajima2018uniaxial,okamura2017emergence,ukleev2020metastable,brearton2023observation}. Furthermore, polishing can impose both uniaxial and torsional strain on a crystal’s surface, thereby modifying the chirality of helical magnetic textures~\cite{fedorov1997interaction,tarnavich2017magnetic}. We acknowledge that such mechanical strain from sample preparation, combined with translational symmetry breaking and anisotropic exchange interactions at the sample surface, likely contribute to SSS stabilisation. However, at present, we cannot unambiguously distinguish between these extrinsic and intrinsic contributions.
	
	Based on our observations, we suggest that stripe and fan states are plausible candidates that explain the exotic double-periodic square-like texture observed in our experiments on \ce{Cu2OSeO3}. However, the alternating chirality of domain walls contradicts the single chiral character of all the other modulations in \coso~stabilised by the DMI. Obviously, the CD-REXS has the same sign for CSL and SSS modulations (Fig. \ref{fig3}\textbf{e}), while for the fan phase it would be simply absent due to the degenerate chirality. We conclude that the stripe texture, which preserves the chiral symmetry given by DMI in the presence of surface demagnetisation fields, is a more likely candidate to explain the additional phase.
	
	Theoretical description of the SSS emergence and its coexistence with the CSL in the narrow field range is a challenging problem. A continuum of metastable helical (conical) states differing by the wave number has been theoretically predicted for uniaxial helimagnets \cite{laliena2023continuum}. However, in the absence of a microscopic theory including multiple anisotropic interactions, to address the basic properties of the SSS phase, we can use a phenomenology of the CSL problem. We note that the strong third harmonic peak from SSS observed experimentally indicates strong axial (second order) anisotropy of a surface nature. Moreover, we can expect that the DMI strength is not much affected by the surface, so the corresponding length scale $1/Q$ is intact. Under these assumptions, one can estimate the required anisotropy strength to have the structure with the doubled period and consequently calculate the third harmonic amplitude (similar to Fig.~\ref{fig3}\textbf{a}-\textbf{d}). Importantly, these calculations show a reasonable agreement with the experimental data, which provides additional support for our claim about the strongly modulated helicoid, approaching the stripe domain nature of the SSS structure.
	
	Finally, the phase diagram for SSS is presented in Fig. \ref{fig4}. The narrow phase region indicates a balance of interactions: restricted magnetic field range, specific angles of the required magnetic field application, and lower temperatures required to stabilise this phase, compared to helical or conical phases. Interestingly, the new phase emerges in the same temperature range where exotic tilted conical and disordered skyrmion phases are stable \cite{crisanti2023tilted}, suggesting that the surface-modified anisotropies responsible for SSS formation may be intimately connected to the bulk magnetocrystalline anisotropy that governs the low-temperature phase diagram of \ce{Cu2OSeO3}. The emergence of the new surface state challenges existing theories of cubic chiral magnets and calls for further theoretical and experimental efforts to understand the complex landscape of magnetic interactions in \coso. Our results further warrant careful inspection of similar surface-confined spiral states in other chiral skyrmion hosts, such as Co-Zn-Mn alloys, which host potential for possible room-temperature applications.
	
	\section{Methods}
	
	\noindent\textbf{\textit{Sample preparation}}
	
	\noindent The \coso\, single crystals employed in resonant elastic X-ray scattering (REXS) and small-angle neutron scattering (SANS) experiments were produced via the chemical vapor transport (CVT) method, as detailed in the Ref.~\cite{baral2022tuning}. One single crystal was aligned along the [001] direction, and the in-plane \{100\} and \{110\} domains were also identified. In the subsequent step, the cross-section of the single crystal was hand-polished. Finally, the (001) facet was cut and polished to create the desired disc-shaped crystal. All key results presented in this manuscript, including both REXS and SANS data, were obtained on this crystal. The (001) facet of the second single crystal was intentionally kept rectangular. After polishing, the alignment of both single crystals was verified using a Laue diffractometer.
	
	\noindent\textbf{\textit{Resonant elastic X-ray scattering}}
	
	\noindent \textit{REXS experiment}: The REXS experiment was carried out using the MaReS chamber at BL-29 BOREAS at ALBA synchrotron light source (Barcelona, Spain). Circularly polarised soft x-rays with energy $E=930.8$\,eV at the Cu $L_3$ edge were used to maximise the intensity of magnetic scattering. The combination of the incidence angle $\theta=48^\circ$ corresponding to the Bragg condition and the energy of x-rays allows to estimate the penetration depth of 50\,nm from the sample surface \cite{zhang2018reciprocal}. The (001)-polished disc-shaped \coso~crystal was mounted on an azimuthally-rotatable sample holder to establish the relative alignment between the crystal axes and the in-plane magnetic field. With [001] axis being out of the sample plane, the remaining two $\langle100\rangle$ and two $\langle110\rangle$ axes were in the sample plane. The sample was zero-field cooled down to base temperature of 18\,K by the He-flow cryostat. The in-plane magnetic field up to 2\,T was provided by the high-temperature superconducting (HTS) magnet (HTS-110, New Zealand). Measurements were carried out using the zero-field warming protocol from the base to the target temperature. Field scans were performed using field ramping down cycles, always starting from the saturation field of 100\,mT. The REXS intensity was detected by the CCD detector $2048\times2048$ pixels (XCAM, UK). The total acquisition time of each REXS pattern was from 1 to 5\, seconds depending on the magnetic state of the sample.
	
	\noindent \textit{SAXS experiment}: SAXS experiment was carried out on a \coso~lamella in forward geometry using VEKMAG endstation at the BESSY-II synchrotron (Berlin, Germany). Details of the experiment are given in Ref.~\cite{baral2023direct}. Detailed magnetic field scan in SAXS geometry at $T=14$\,K, with the same measurement protocol as the one discussed in the main text, does not reveal any signature of the SSS state -- the $Q_{\mathrm{SSS}}$ peak is absent in the whole field range (Supplementary Materials Fig.~S4).
	
	\noindent\textbf{\textit{Small-angle neutron scattering}}
	
	\noindent Small-angle neutron scattering (SANS) experiments were carried out at the ILL using the D33 instrument \cite{dewhurst2016small} and at PSI using the SANS-I instrument with a neutron wavelength of 10\,\AA~and the sample-detector and collimator-sample distances of 12.8~m. The same \coso~single crystal was probed as in the REXS measurement at ALBA. Two configurations were used: $\mu_0 H|| [110]$ (Figs.~S5\textbf{a},\textbf{b}) and $\mu_0 H|| [110] + 8^\circ$ (Fig.~S5\textbf{c},\textbf{d}) in the field-transverse to the beam geometry. At each set point, the magnetic field was stabilised before collecting data. We employed the so-called rocking curve measurement, where the magnet+sample assembly was rotated around the Bragg condition in a small angular range. This method enables us to access the total intensity of the Friedel pairs. SANS data was analysed using GRASP software \cite{dewhurst2023graphical}.
	
	\section{ACKNOWLEDGEMENTS}
	
	We sincerely thank A. Leonov for fruitful discussions. REXS measurements were carried out at the beamline BL-29 BOREAS at ALBA synchrotron as a part of proposals 2021024923 and 2023027505. Transmission-REXS measurement was carried out at the beamline PM-2 VEKMAG at BESSY II synchrotron as a part of the proposal 212-10682. SANS measurements at the D33 instrument (Institut Laue-Langevin) were performed as a part of the proposal 5-41-1226. P.R.B. acknowledges Swiss National Science Foundation (SNSF) Postdoc.Mobility grant P500PT\_217697 and Return CH Postdoc.Mobility grant P5R5-2\_239356 for financial assistance. P.R.B., J.S.W., A.M., V.U. acknowledge funding from the SNSF Projects Sinergia CRSII5\_171003 NanoSkyrmionics. J.S.W., V.U., and P.R.B. also acknowledge SNSF Project 200021\_188707, SNSF National Centers of Competence in Research in Molecular Ultrafast Science and Technology (NCCR MUST-No. 51NF40-183615) and SNSF grant no. 200020\_182536 (Frustration in structures and dynamics), respectively. V.U., C.L., F.R. acknowledge financial support by the German Federal Ministry for Education and Research (BMBF project No. 05K19W061) and financial support by the German Research Foundation via Project No. SPP2137/RA 3570. O. I. U. acknowledges financial support from the Institute for Basic Science (IBS) in the Republic of Korea through Project No. IBS-R024-D1. M.T.L. acknowledges the financial support of the Science and Technology Facilities Council (STFC) and the ISIS Neutron and Muon Source.
	
	\section{DATA AVAILIBILITY}
	
	\noindent All experimental data presented in the figures that support the findings of this study are available at the Zenodo repository link: \href{https://doi.org/10.5281/zenodo.19700525}{https://doi.org/10.5281/zenodo.19700525}.
	
	\section{AUTHOR CONTRIBUTIONS}
	
	\noindent P.R.B. and A.M. prepared and characterised the samples; P.R.B., S.H.M., P.G., M.V., C.L., F.R. and V.U. performed x-ray scattering measurements; P.R.B., M.T.L, R.C.,  N-J.S., J.S.W., and V.U. performed the SANS experiments. P.R.B. and V.U. analysed the data; O.U. developed the theoretical expressions based on a perturbative approach. P.R.B., O.U. and V.U. wrote the manuscript; P.R.B., J.S.W., and V.U. jointly conceived the project.
	
	
	\section{COMPETING INTEREST}
	
	\noindent The authors declare no competing interests.
	
	\section{References}
	
	\bibliography{biblio}

	\onecolumngrid
	
	
\end{document}